\documentclass[11pt,a4paper]{article}
\usepackage{amsfonts,amsmath,amstext,amssymb}
\usepackage{dsfont}
\usepackage{theorem}    
\usepackage{a4wide}

\usepackage{color}

\newtheorem{lema}{Lemma}[section]

\newtheorem{rema}[lema]{\bf Remark}

\theorembodyfont{\rmfamily}

\usepackage{graphicx}

\title{The Friedmann cosmological models revisited as an harmonic motion and new exact solutions}

\author{\footnote{The authors are partially supported by the
Spanish MEC-FEDER Grant MTM2010-18099. }
Rafael M. Rubio and  Juan J. Salamanca\\[6mm]
 Departamento de Matem\'aticas, Campus de Rabanales, \\[0.5mm]
Universidad de C\'ordoba, 14071 C\'ordoba, Spain,\\[0.5mm]
E-mails\textup{:\texttt{\;rmrubio@uco.es},\,\,\texttt{jjsalamanca@uco.es}}}

\date{}

\begin{document}

\maketitle

\begin{abstract}
A new approach for arbitrary dimension to the Friedmann cosmological models is presented. Taking suitable changes of the parameters of the spacetime the harmonic motion equations appear, where the curvature determines the angular frequency. Some physical interpretations are also explained. As a consequence of our approach, we give new exact solutions to the Einstein's equation when the observer is not comoving with the perfect fluid, as well as, when the pressure not vanishes, including a new procedure to acquire exact solutions for cases of universes at the dark energy dominated stage.

\end{abstract}

\thispagestyle{empty}

\vspace{1,5mm}



\hyphenation{ma-ni-fold}

\section{Introduction}

According to the \emph{cosmological principle} (see, for instance, \cite[Chapter V]{CB}), the physical 
space should be looked isotropic and homogeneous, at least
for appropriate large scales. This physical condition
can be geometrically interpreted considering the restspace  of certain observer field as a space form (a Riemannian manifold with constant sectional curvature).
On the other hand,  it seems reasonable not to take into account any local physical objects, such as
galaxies.
All this leads to consider Robertson-Walker spacetimes as good candidates to obtain exact solutions of
the Einstein field equation (see \cite[Cap. 12]{ON})

$${\rm Ric}-\frac{1}{2}Sg=8\pi T.$$

Currently, the interest of General Relativity in arbitrary dimension is notable (see \cite{Sha} and refrences therein) for several reasons as the creation of unified theories, understanding of general features from toy cosmological models ((2+1)-dimension), etc.
 
For a $(n+1)$-dimensional {Robertson-Walker (RW)} spacetime  we mean a product manifold $I \times F$,
of an open interval $I$ of the real line $\mathbb{R}$ and an $n(\geq 2)$-dimensional (connected) Riemannian
manifold $(F, g_{_F})$ of constant sectional curvature, endowed with the Lorentzian metric
\begin{equation} \label{metricaRW}
 g=\pi_{_I}^ *(-dt^ 2)+f(\pi)^ 2\pi_{_F}^ *(g_{_F})\equiv-dt^2 + f(t)^2 \, g_{_F} ,
\end{equation} where $f$ is a positive smooth function on $I$, called warping function, and $\pi_{_I}$ and $\pi_{_F}$ denote the projections onto $I$ and $F$ respectively. We will denote this Lorentzian manifold by 
$\overline{M} = I \times_f F$. The $(n+1)$-dimensional spacetime $\overline{M}$ is a warped product,
in the sense of \cite[C. 7]{ON}, with base $(I,-dt^2)$, fiber $(F,g_{_F})$ and warping function $f$.

\vspace{3mm}
On the other hand, recall that a \emph{perfect fluid} (see, for example, \cite[Def. 12.4]{ON}) on a 
spacetime $\overline{N}$ is a triple $(U,\rho, p)$ where
\begin{enumerate}
 \item $U$ is a timelike future-pointing unit vector field on $\overline{N}$ called the \emph{flow
 vector field}.
 \item $\rho, p \in C^\infty (\overline{N})$ are, respectively, the \emph{energy density function} and the 
 \emph{pressure function}.
 \item The stress-energy tensor is 
 $$
 T= (\rho + p ) \, U^* \otimes U^* + p \, {g} ,
 $$ where ${g}$ is the metric of the spacetime $\overline{N}$.
\end{enumerate}

As it is usual, we take the flow given by the vector field $U:=\frac{\partial}{\partial t}$, which serves to represent a perfect fluid in a RW spacetime.
On the other hand, it is also natural in this context to assume that the perfect fluid is dust, i.e., the pressure $p=0$, and as a direct consequence the 
 stress-energy tensor is given by
$$
T_{_{dust}} = \rho \,  \, U^* \otimes U^* \, .$$

When the RW spacetime has dimension 4 and it is exact solution to Eintein's equation as a dust with flow vector field $\frac{\partial}{\partial t}$, it is called a Friedmann cosmological model.

Note that the family of observers $U=\frac{\partial}{\partial t}$ are comoving with the dust.

In this work we obtain the Friedmann cosmological solutions to the Einstein's equation for arbitrary dimension. Our intrinsic approach (without taking coordinates on the fiber) is based in the definition of new geometric elements for the $(n+1)$-dimensional spacetime, which to allow one to obtain the exact solutions (see Section 4) with quite simple computations (compare with \cite{Sha}). In addition and to an important degree, this new approach will allow to obtain spacetime models when the observer field is not comoving with the flow vector field (see Section 6), as well as, new exact solutions in cases with pressure $p\not=0$, including universes models with negative presure (see section  7), which can be suitables to describe accelerated expanding universes (dark energy dominated stage). On the other hand, the intrinsic geometrical approach is subject to be applied to other class of warped product spacetimes (see section 8).

\section{Structural equations of a perfect fluid in Robertson-Walker spacetimes}

A RW spacetime $(\overline{M}=I\times_f F,{g})$, which is a perfect fluid $(\frac{\partial}{\partial_t},\rho,p)$ obeying the Einstein's equation, must
 satisfy,

\begin{equation} \label{rhofluido}
8 \pi \rho = \frac{n(n-1)}{2} \frac{k}{f^2} + \frac{n(n-1)}{2} \frac{f'^2}{f^2}
\end{equation} 
\begin{equation} \label{pfluido}
-8 \pi p = \frac{(n-1)(n-2)}{2} \frac{k}{f^2} + \frac{(n-1)(n-2)}{2} \frac{f'^2}{f^2} + (n-1) \frac{f''}{f} \, ,
\end{equation}

\noindent being $n$ the dimension of its fiber, $f$ the warping function and $k$ the constant sectional curvature of $F$ (see \cite[Cap. 12]{ON}.

\begin{rema}
Observe that from previous equations the case $n=1$ entails $\rho=p=0$, and consequently it can only model a vacuum spacetime.
\end{rema}

Standard computations show (see, for instance, \cite[Cor. 12.13]{ON}) that the following equation holds, 
\begin{equation}\label{derivada}
d\rho(\frac{\partial}{\partial t})= -n(\rho + p) \frac{f'}{f} \, .
\end{equation}
\noindent When the perfect fuid is dust, from  the previous equation it is easy to see,
\begin{equation}\label{M}
8 \pi \rho f^n = M,
\end{equation}

 \noindent where $M$ is a positive constant.

\subsection{Solution in $3$-dimensional spacetimes}

Because it should be treated as a special case, in this section we show the solution for $3$-dimensional spacetimes when the perfect fluid is dust. 
Equations (\ref{rhofluido}) and (\ref{pfluido}) are written for this case as,
$$
M= k +f'^2 ,
$$ where $8 \pi \rho f^2=M$ is a constant. This equation is easily solved, and its solutions fits into the following classes,

\begin{enumerate}
	\item When $M>k$, expanding universes, corresponding to $f(t) = \sqrt{M-k} \, (t+t_0)$. It has an initial singularity in $t=t_0$. Observe that
	it can be taken a temporal origin in $t_0=0$.
	\item When $M>k$, contracting universes, corresponding to $f(t) =  \sqrt{M-k} \, (t_f-t)$. It has a final singularity in $t=t_f$. Observe that
	it can be taken a temporal final in $t_f=0$. A time-orientation reversion fits into the previous case.
	\item When $M=k$, static universes are obtained, i.e., $f(t) =A$, $A \in \mathbb{R}^+$. 
\end{enumerate}

\section{Structural equations in a distinguished coordinate system}

From now on, we assume that the fiber of the RW spacetime has dimension $n \geq 3$.
First we define the two principal elements for the description of our study.

Let us define the following function
$$
X:= f^{(n-2)/2} \, .
$$ Also, we define the \emph{universal angle} 
$$
\theta := \int_{t_0}^t \frac{ds}{f(s)} \, ,
$$ where $t_0 \in I$ an  arbitrary  point. 
The timelike coordinates $t$ and $\theta$ satisfy the following differential equation,
\begin{equation} \label{tetat}
d\theta = \frac{dt}{f(t)} \, .
\end{equation}
Some computations show that equations (\ref{rhofluido}) and (\ref{pfluido}) are rewritten as

\begin{equation} \label{e1}
\frac{16 \pi}{n(n-1)} \, \rho \, X^{2n/(n-2)} = k \, X^2 + \frac{4}{(n-2)^2} \left(  \frac{dX}{d\theta} \right)^2
\end{equation} 
\begin{equation} \label{e2}
\frac{d^2 X}{d\theta^2} = - \frac{(n-2)^2 \,k}{4} \, X - \frac{n-2}{n-1} \,4 \pi \, p \,  X^{(n+2)/(n-2)} \, .
\end{equation} 

\begin{rema} \label{metricaXtheta}
It should be noticed that the metric of the spacetime, in relation to $X$ and $\theta$, takes the following form,
\begin{equation}\label{conformal}
g=   X^{4/(n-2)}(\theta) \left\{ -d\theta^2 +g_{_F} \right\} \, 
\end{equation}
\noindent being then conformally related with a static spacetime.
 \end{rema}

\section{Dust as an harmonic motion}

In the case of dust, equations (\ref{e1}) and (\ref{e2}) are written as follows,
\begin{equation} \label{final1}
M= k \, X^2+ \frac{4}{(n-2)^2} \left(  \frac{dX}{d\theta} \right)^2
\end{equation} \begin{equation} \label{final2}
\frac{d^2 X}{d\theta^2} = - \frac{(n-2)^2 \, k}{4} X ,
\end{equation} where $M:=16 \pi \rho f^n /(n(n-1))$ is a constant, as corresponding to dust.
When $k \neq 0$, the equation (\ref{final2}) represents mathematically 
the equation of an harmonic simple oscillator, and the previous one its first integral, corresponding
to its energy.
When $k=0$, it seems to represent mathematically a free particle.

The solutions can be easily obtained as follows (compare with \cite[C.12, Rem. 21]{ON} and \cite{Sha}; see also
\cite{Hawking}).

\subsection{Flat fiber (the case when $k=0$)}

In this spetial case, the equation (\ref{final1}) reads,
$$
\frac{dX}{d\theta} = \frac{(n-2)}{2} \sqrt{M} \, .
$$  Taking $\theta_0 =0$, we obtain
\begin{equation} \label{flatX}
X= \frac{(n-2)}{2} \sqrt{M} \,  \theta \, . 
\end{equation} With this computations (and in general), it is obtained the parametrized solution of the warping function; that is, $f=f(\theta)$, $t=t(\theta)$.
So, it can be written,
$$
f(\theta) =  \left(  \frac{(n-2)}{2} \sqrt{M} \right)^{2/(n-2)} \theta^{2/(n-2)} \, .
$$ And the dependence with $t$ is followed from (\ref{tetat}),
$$
t= \int_{\theta_0}^\theta f(s) ds =  \int_{\theta_0}^\theta  \left(  \frac{(n-2)}{2} \sqrt{M} \right)^{2/(n-2)} s^{2/(n-2)} ds \, .
$$ After some computations, it can be expressed $f$ as a function of $t$,
$$
f(t) = \left( \frac{n \, \sqrt{M}}{2} \, t  \right)^{2/n} \,  .
$$

\begin{rema}
Observe that a time-reversal solution is obtained by a change of sign of $\theta$. On the other hand, note that $X$ presents the same
behaviour than the case of $2$-dimensional fiber.
\end{rema}

\subsection{Fiber with positive constant sectional curvature ($k>0$)}

The solution is easily followed in terms of $X$ and $\theta$ and it corresponds exactly to those of an harmonic motion,
\begin{equation} \label{Xkp}
X(\theta) = \sqrt{\frac{M}{k}} \, \sin \left(  \frac{(n-2)}{2} \sqrt{k} \, \theta \right) \, .
\end{equation} This harmonic motion has angular frequency

$$
w := \frac{(n-2)}{2} \sqrt{k} \, .
$$ This is the reason why $\theta$ has been called universal angle. In this sense, $w$ could represent a \emph{universal angular frequency}.
Observe that it is related to the constant sectional curvature of the fiber. These equations could lead to think in the dust as a monochromatic wave.

\ 

Following the same procedure than the  previous case, it is possible to give explicitely the solution (the origin is taken in the initial singularity),
$$
f(\theta) = \left( \frac{M}{k} \right)^{1/(n-2)} \, \sin^{2/(n-2)} (w \, \theta)
$$ $$
t= \left( \frac{M}{k} \right)^{1/(n-2)} \int_0^\theta \sin^{2/(n-2)} (w \, s) \, ds \, .
$$

\begin{rema}
When $n=3$, the integral on $s$ can be expressed in terms of simple functions, and it can be showed that it coincides with \cite[C.12, Rem.21]{ON} when $k=1$.
For  dimension $n>3$, the integral is expected not to be expressed in terms of simple functions.
\end{rema}

\begin{rema} (See \cite[C.12, def. 16]{ON})
Put $t=0$ as the initial singularity and $\theta_f=2\pi/((n-2) \sqrt{k})$ as the final singularity corresponding to a time coordinate value $t_f$. Then,
 it is not difficult to see that
the initial singularity is a Big-Bang, i.e.,
$$
\lim_{t\rightarrow 0^+} f =0 
$$ $$
\lim_{t \rightarrow 0^+}  f' = +\infty \, ,
$$ and the final singularity is a Big-Crunch, i.e.,
$$
\lim_{t\rightarrow t_f} f =0 
$$ $$
\lim_{t \rightarrow t_f}  f' = -\infty \, . $$
\end{rema}

\subsection{Fiber with negative constant sectional curvature $k<0$}

In this case, equation (\ref{final1}) reads,
$$
M= -|k| X^2 + \frac{4}{(n-2)^2} \left( \frac{dX}{d\theta} \right)^2 .
$$ It is known that its solutions are given by $X= a \, \sinh(w \, \theta)$, for accurate $a,w \in \mathbb{R}^+$, but we will derive this in another way.
The previous equation can be expressed as
$$
-M = |k| X^2 + \frac{4}{(n-2)^2} \left( \frac{dX}{d(-i \theta)} \right)^2 ,
$$where $i$ stands for the imaginary unit. Then, we can think in these solutions as the associated for the case $k>0$ with the 
following changes: $M\mapsto -M$ and $\theta \mapsto -i\theta$. The sign minus here is taken in order to preserve the time-orientability.
With this in mind, the solution for a fixed $k<0$ is the same than $|k|$ with the opportunes changes. It can be done explicitly and it is obtained,
$$
X= \sqrt{ \frac{M}{|k|} } \sinh \left( \frac{(n-2)}{2} \sqrt{|k|} \, \theta \right)   \, .
$$ The solution in terms of  $f$ and $t$ is computed to be
$$
f(\theta) = \left( \frac{M}{|k|} \right)^{1/(n-2)} \, \sinh^{2/(n-2)} (w \theta)
$$ $$
t= \left( \frac{M}{k} \right)^{1/(n-2)} \int_0^\theta \sinh^{2/(n-2)} (w s) \, ds \, .
$$

\begin{rema}
In this case, $\theta \in \left( 0,\infty \right)$ and it does not present a final singularity. But it can be
proved that the initial singularity is a Big-Bang.
\end{rema}

Before making some physical considerations of these facts, we will derive the solution for $k=0$ as a certain limit of spacetimes \cite{Geroch}.

\subsection{Flat fiber as certain limit}

We will show that the solution for $k=0$ can be recovered from the solutions for $k>0$ and $k<0$ taking an appropriate limit of spacetimes (see \cite{Geroch}).
For $k\rightarrow 0^+$,
$$
X_{k=0} = \lim_{k\rightarrow 0^+} \sqrt{  \frac{M}{k} } \sin \left( \frac{(n-2)}{2} \sqrt{k} \theta \right) = \frac{(n-2)}{2} \sqrt{M} \theta .
$$ That is, the solution for $k=0$ (see section 4.1). 

The same computations can be done for the case $k\rightarrow 0^-$ and it can be checked to obtain the same solution.

\ 

Observe that the case of flat fiber corresponds to a vanishing universal angular frequency.
The case $k \rightarrow +\infty$ can be understood to be $X$ vanishing, and the universe has no evolution, that is, it has no
temporal development. This could be in agreement with the physical experience of falling into a black hole. The case $k \rightarrow -\infty$
can be interpreted as a blow-up of the universe, taking infinite energy. This is not expected to occur from a physical point of view.

\begin{rema}
Note that the equations of an harmonic motion appears only when $n \geq 3$. It should be also noted that the
angular universal frequency depends on the dimension. A notable question in  physics is to explain why the dimension
of the physical space is three, at least at accurate scales. From our point of view, note that the dimension three is the minimum dimension in which the spacetime dust is written as an harmonic motion.
We could even think that before the Big-Bang singularity all were reduced to certain quantity of energy and after, 
 the nature fells into the least dimension possible; this could be reasonable if we think that the dimensions built require energy.
\end{rema}

\section{A space of spacetimes}

Arising from the previous considerations, we can represent all the solutions of dust spacetimes (fixed 
an arbitrary dimension $n \geq3$) as  points $(M,k)\in\mathbb{R}^2$ with $M>0$.

On the other hand,  when $(M,k)$ is a point of the semiplane $M<0$, we associate the related dust solution, obtained when the mathematical formalism used, takes the metric tensor of the spacetime with signature (1,-1,-1,-1).

 Only for mathematical sakeness, we
define the upper semiplane $M > 0$ as the \emph{space of universes}, and the lower semiplane $M < 0$
as the \emph{space of antiuniverses}. There exists a natural association
between a universe $u$ and an antiuniverse $u^{*}$ given by 
$$
u=(k,M) \longleftrightarrow u^* = (-k,-M) .
$$ Observe that this association preserves the physical {\it mass-factor} $|M|$ and its geometrical (and physical) structure. It would be interesting a comparison with \cite{Stenger}.

\ 

This formal artifice suggests to think about a pair universe-antiuniverse created simultaneously in the Big-Bang, but so that the total
mass is zero. In order to do that, we could think the mass of the antiuniverse as a negative mass but with identical absolute value  to that of its associated universe. Then,
the sum vanishes. The {\it factor-mass} will be determined according to the energy existent before Big-Bang. It should be noticed that this process could be in accordance with the physical problems of
creating particles-antiparticles from vacuum. 

The reason to propose this idea follows from the problem of the asymmetry matter over antimatter (see \cite[Section 12.3]{G}). It is normally assumed 
that the Big-Bang created
matter and antimatter in exactly equal amounts. As it is known, if matter and anti-matter (for example, an electron and a positron) 
encounters, they annihilate each other, and energy is liberated in the process. But this doesn't explain the preponderance of
leftover matter. The most likely possibilities are:

\begin{itemize}
	\item This is a local phenomenon; that is, somewhere else out there there are antimatter
regions. Therefore the ratio of matter and antimatter is balanced. However, there
is no evidence for this. Another consideration is present is this case: what kind of
mechanism would produce this ``local difference"?
\item A violation of the symmetry of the laws of Quantum Mechanics, concretely, the
Charge-Parity-violation. The central problem here is that this assumption is
showed to be not well understood.
\end{itemize}

The following idea could give a hint to solve this. Assume that a pair universe-antiuniverse was created simultaneously in the Big-Bang process.
Then, it seems natural to think that they two interacted until certain instant in which both was separated. Each universe could create exactly the
same matter and antimatter amounts, but, after the separation, each part could have an overall of one kind of matter over the other. Consequently,
if one universe presents an overall of matter than antimatter, its associated antiuniverse presents the opposite case, and the pointed problem
could be solved without change basic laws of nature. With this conjeture, is there any possibility to make any scenario to produce the observed value of the baryon asymmetry, namely, the baryon-to-photon ratio (balance matter-antimatter) (see \cite{Ko}).

\section{A new solution to moving dust}

Suppose that a determined observer see its universe moving with a constant speed (see \cite{DF}) and the only substance that he can measure is dust-molecules in motion respect to him.

\begin{quote}{\it What will be his proposal of spacetime, if its restspace is assumed to be $\mathbb{R}^n$?}
\end{quote}

 We propose a family
of spacetimes to model that physical scenario. Hence, let $V \in (-1,1)$ be a constant, and $\partial_x$ a coordinate
vector field on $\mathbb{R}^n$. Consider the product manifold $I \times \mathbb{R}^n$, being $I \subset \mathbb{R}$ an open interval, endowed with the
following Lorentzian metric,
\begin{equation} \label{llanomueve}
g_{_{V \, \mathrm{flat}}}= \xi \left( \theta - V x \right)^{4/(n-2)}  \left\{ -d\theta^2 +dx^2+ \sum_{i=2,\ldots,n} dx_i^2 \right\} ,
\end{equation} where, for manageability, it has been defined,
$$
\xi = \left( \frac{(n-2) \sqrt{M} }{2} \right)^{4/(n-2)} \frac{1}{(1-V^2)^{2/(n-2)}} \, .
$$ Observe that this family of spacetimes extends properly the Friedmann model with flat fiber. 
 Let computate the stress-energy tensor, $T_{_V}$, of this spacetime. To make this computation, consider the following change of
coordinates
$$ 
{ \theta' =  \frac{\theta+V x}{\sqrt{1-V^2}} 
\atop
x'  = \frac{x+V \theta}{\sqrt{1-V^2}} \, .
}   $$  It can be showed that $g_{_V}$ in the prima coordinates takes 
the form (\ref{conformal}). Then, 
$$
T_{_V} = \rho d\theta' \otimes d\theta ' =  \frac{\rho}{1-V^2}  ( d\theta+ V d x  ) \otimes ( d\theta+ V d x  ) .$$
\noindent Therefore, the Lorentzian manifold $(I \times \mathbb{R}^n, g_{_{V \, \mathrm{flat}}})$ is a exact solution to the Einstein's equations  with energy-tensor $T_V$.

 This family of spacetimes should be interpreted as a dust fluid with vector field flow 
$$
U = \frac{1}{  \xi^{1/2} (\theta -V x)^{2/(n-2)} \sqrt{1-V^2}}  \left\{  \partial_\theta + V \partial_x  \right\} \, .
$$  
For every event $p$ in a spacetime of this family, there exists two distinguished observer fields, $O:=\partial_\theta/g(\partial_\theta,\partial_\theta)^{1/2}$ and
$U$. Note that both observer fields cover the whole spacetime. Then, it can
be computed the following: considering $O (p)$ and $U(p)$ in the tangent space in $p$, then the
hyperbolic angle between these two tangent vectors is $\cosh \alpha (p)= -g(O,\partial_\theta)(p)$. Hence, $\cosh \alpha = 1/\sqrt{1-V^2}$
and the relative speed that $O$ measures from $U$ is a constant value $V$. This should be interpreted as
the speed of the dust that measures the observer in  $O$ is constant (compare with \cite{K}).

On the other hand, each integral curve of $O$ represents an observer in $O$, and from the previous
considerations, he must declare that the velocity of the dust fluid that he measure in each point of its
worldline is constant.

Emerging on the expression of $T_{_V}$, there exists another reasons to indicate that this is the expected spacetimes. First, observe
that it appears a non-vanishing term $T_{_V}(\partial_x, \partial_x)$, that should be interpreted as the energy from the motion of that direction.
That is, as the momentum associated that the motion of the dust fluid has in its spatial component. It is natural that the stress-energy tensor has
the physical information of that component, and it seems also natural to have the form expressed before.

\ 

Analogously, it is proposed a family of spacetimes when the restspaces  are assumed to be the hyperbolic space, $\mathbb{H}^n(k)$, i.e.,
a Riemannian manifold with constant sectional curvature $-k$ (and simply-connected). Observe first that it is required a
vector field on $\mathbb{H}^n(k)$ whose integral curves are geodesic, since they serve for the local description on the physical space of the
motion of the dust. In fact, observe that $\partial_x$ obeys that 
condition when $\mathbb{H}^n(k)$ is described by $\mathbb{R}\times \mathbb{R}^{n-1}$
with metric
$$
g_{_{\mathbb{H}^n (k)}} = dx^2 + e^{- \sqrt{|k|} x} \, g_{_{\mathbb{R}^{n-1}}} .
$$ Therefore, for this case, we could write
\begin{equation} \label{hiperbolicomueve}
g_{_V} = \left(\sqrt{\frac{M}{|k|}} \sinh \left(  \frac{(n-2) \sqrt{|k|}}{2} \frac{\theta + V x}{\sqrt{1-V^2}}  \right)\right)^{4/(n-2)} 
\left\{  -d\theta^2 +dx^2 + e^{- \sqrt{|k|}(x+V \theta)/\sqrt{1-V^2}} g_{_{\mathbb{R}^{n-1}}} \right\} .
\end{equation}
The same arguments previously done for the flat case can be analogously made in this case, and this indicates
its plausibility for this problem.

\ 

Observe that the family of spacetimes given by (\ref{hiperbolicomueve}) extends properly the family of spacetimes dust with non-positive curvature of the fiber.

\ 

The study of this family of spacetimes would be motivated from some arguments. On one hand,
observe that a total description of a spacetime of this family is encoded by a type wave function. Apart
from its interesting and simple form, it could serve to study some cosmological models of universe.

On the other hand, in a natural way, it appears a direction in the physical space. Observe that no
fundamental theoretical basis is modified (such as homogeneity of the dust fluid). This family of
spacetimes would be of interest to study the dark flow, recently discovered (see, for instance, \cite{DF}).

\

\section{Non-vanishing pressure}

It is normally argued a linear relationship between $\rho$ and $p$ in the form $p = \beta \, \rho$, i.e., universes described by a barotropic perfect fluid (see \cite[section IIB]{CST}). Therefore, the ``motion equation''
for this case can be showed to be

\begin{equation} 
\frac{d^2 X}{d\theta^2} = - \frac{k \, (n-2) \, (n-2+\beta \, n)}{4} \, X - \frac{n\, \beta}{n-2} \frac{1}{X} \, \left( \frac{dX}{d\theta} \right)^2  \, . 
\end{equation}

\noindent As it is well-known, this second order ordinary differential equation has assured local existence and uniqueness of solutions. Moreover, for the case of flat fiber (observations have shown that the current universe is very close to a spatially flat geometry \cite[section II] {CST}), some solutions $X=X(\theta)$ are given by $X= A \, \theta^B$, for convenient
constants $A, B$.
Note also that the general behaviour is expected to not depend strongly on the dimension when $n\geq 3$. In particular, this produce can also be applied to an universe at the dark energy dominated stage, i.e., the dynamical equation of state parameter $\beta$ defined by the pressure of dark energy over its density energy $\beta=p/\rho$. Hence we can obtain exact solutions $X(\theta)$ for these last scenarios (compare with the very interesting references on modified gravity and dark energy \cite{B}, \cite{NO} and \cite{N}).

Another relationships between $\rho$ and $p$ can also be considered and the equation of that motion could then be stated.

Finally, to end this paper, we will consider some new perfect fluid solutions to the Einstein's field equation, whose pressure is not zero. 

Note that in Robertson-Walker spacetime the area element for each surface $S$ in a slice $t=t_0$ is proportional with factor $X^{4/(n-2)}(\theta_0)$ to the respective area element of the surface $\pi(S)$ in the fiber, where $\pi$ denotes the natural projection on the second factor. Then, we study the case in which the pressure of the perfect fluid satisfies $p X^{4/(n-2)} = p_{_0}$, where $p_{_0} \in \mathbb{R}$
is a  constant. Note that the absolute value of pressure is increasing (resp. decreasing) when the universe is contracting (resp. expanding). 
This case is very easy to solve, since its solutions can be expressed as spacetimes in the form of Remark 3.1. In fact, from (\ref{e2}),
it is solved for $k \geq 0$, obtaining
$$
X(\theta) = A \, \sin (w_{_+} \theta) \, , 
$$ where $A \in \mathbb{R}$ is a positive constant, and
$$
w_{_+}^2 = \frac{(n-2)^2 \, k}{4} + \frac{4 (n-2) \pi}{(n-1)} \, p_{_0} \, .
$$ Also, for $k \leq 0$,
$$
X(\theta) = A \, \sinh (w_{_-} \theta) \, , 
$$ where $A \in \mathbb{R}$ is a positive constant, and
$$
w_{_-}^2 = \frac{(n-2)^2 \, k}{4} - \frac{4 (n-2) \pi}{(n-1)} \, p_{_0} \,  .
$$ From (\ref{e1}), observe that $\rho X^{2n/(n-2)}$ is not a constant in these models, and, in the hyperbolic case,
this term is not upper bounded.

\section{Conclusions and final comments}

As a consequence of a new method on Robertson-Walker spacetimes  to obtain Friedmann cosmological models, we give several new exacts solutions to the Einstein's equations on different physical scenarios:

a) Perfect fluids dust, whose flow vector field is not comoving with the observer field.

b) Barotropic perfect fluids with equation of state $\beta=p/q$, including the cases of spacetimes at the dark energy dominate stage.

c) Another models with pressure $p\not=0$, being this in functional relation with the area element of certain distinguished spacelike surfaces.

\vspace{2mm}

 On the other hand, the definition of a space of spacetimes allow us to consider a framework for to explain the observed value of the baryon asymmetry.

Finally, note that the Robertson-Walker spacetimes are spatially-homogeneous. To be spatially-homogeneous, which
is reasonable as a first approximation of the large scale structure of the
universe, could not be appropriate when we consider a more accurate
scale \cite{Ra-Sch}. So, it seems reasonable to consider the class of spacetimes called Generalized Robertson-Walker spacetimes (see \cite{A-R-S1}), in which,  our intrinsic approach could be suitable.

\vspace{4mm}
\noindent {\bf Acknowledgements} The authors are grateful to the  referee for 
their deep readings and making suggestions toward the improvement of this article.

\end{document}